\documentclass{article}

    \usepackage[final,nonatbib]{neurips_2022}

\usepackage[utf8]{inputenc} %
\usepackage[T1]{fontenc}    %
\usepackage{hyperref}       %
\usepackage{url}            %
\usepackage{booktabs}       %
\usepackage{amsfonts}       %
\usepackage{lipsum}
\usepackage{caption}

\usepackage[export]{adjustbox}
\usepackage{multirow}
\usepackage{subcaption}
\usepackage{mathtools}
\usepackage{scalerel}
\usepackage{amsmath}       %
\usepackage{nicefrac}       %
\usepackage{microtype}      %
\usepackage{xcolor}         %
\usepackage[doi=false,isbn=false,url=false,eprint=false,style=numeric]{biblatex}
\usepackage{wrapfig}

\newcommand{\grey}[1]{\textcolor{black}{#1}}
\newcommand{\B}[1]{\mathbf{#1}}
\newcommand{\T}[1]{\text{#1}}

\title{Multiscale Metamorphic VAE\\for 3D Brain MRI Synthesis}

\author{%
  Jaivardhan Kapoor%
    \\
  University of Tübingen\\
  \texttt{jaivardhan.kapoor@uni-tuebingen.de} \\
   \And
   Jakob H. Macke\\
  University of Tübingen\\
  Max Planck Institute for Intelligent Systems\\
  \texttt{jakob.macke@uni-tuebingen.de} \\
   \And
   Christian F. Baumgartner\\
  University of Tübingen\\
  \texttt{christian.baumgartner@uni-tuebingen.de} \\
}

\begin{document}

\maketitle

\begin{abstract}
  Generative modeling of 3D brain MRIs presents difficulties in achieving high visual fidelity while ensuring sufficient coverage of the data distribution.
  In this work, we propose to address this challenge with composable, multiscale morphological transformations in a variational autoencoder (VAE) framework.
  These transformations are applied to a chosen reference brain image to generate MRI volumes, equipping the model with strong anatomical inductive biases.
  We show substantial performance improvements in FID while retaining comparable, or superior, reconstruction quality compared to prior work based on VAEs and generative adversarial networks (GANs). %
\end{abstract}
\vspace{-1em}
\section{Introduction}
\label{intro}

The paradigm of generative modeling could be an invaluable tool for better understanding and diagnosing brain disorders.
Deep generative models such as VAEs \cite{vae} and GANs \cite{gan} have enjoyed tremendous success when used for disease attribution maps \cite{vagan}, longitudinal brain aging \cite{dani-net}, and synthetic data generation \cite{equitable, braindiffusion}.
Several previous works \cite{3d-stylegan,adv-ae-3dmri,dani-net, icam-reg,metamorphic,equitable, 3dbraingen} have employed such models to capture the distribution of Magnetic Resonance Images (MRIs) of the brain.
However, due to the complementary strengths of GANs and VAEs, these methods either suffer from blurry outputs, insufficient data distribution coverage, or lack a latent space for further analyses \cite{icam-reg}.

Structural MR images of the brain are morphologically constrained to a smaller set compared to that of natural images such as ImageNet, due to similar brain anatomy across subjects.
Morphologically constrained generative models have been used previously for medical image registration \cite{voxelmorph,augmentationdalca}, where the task is to map an image to a common space for subsequent analysis.
This approach has also been explored previously in \cite{metamorphic,guidedmetamorphic,equitable} for image synthesis.
However, such models have only been shown to work on 2D MRI slices of the brain \cite{metamorphic,guidedmetamorphic}, or do not contain a manipulable latent space for downstream analyses \cite{equitable}.

In this work, we extend the popular VAE approach by taking advantage of the strong anatomical priors in the brain. In contrast to regular VAEs, our approach outputs compositions of \textit{morphological transformations} comprising diffeomorphisms and intensity transformations at different scales. Those are then iteratively applied to a fixed reference MRI template. This allows us to generate high-fidelity MRI volumes while retaining a meaningful latent space.

\section{Methodology}
\label{methodology}
\begin{figure}[t!]
  \centering
  \begin{subfigure}[b]{0.99\textwidth}
    \centering
    \includegraphics[width=0.85\textwidth]{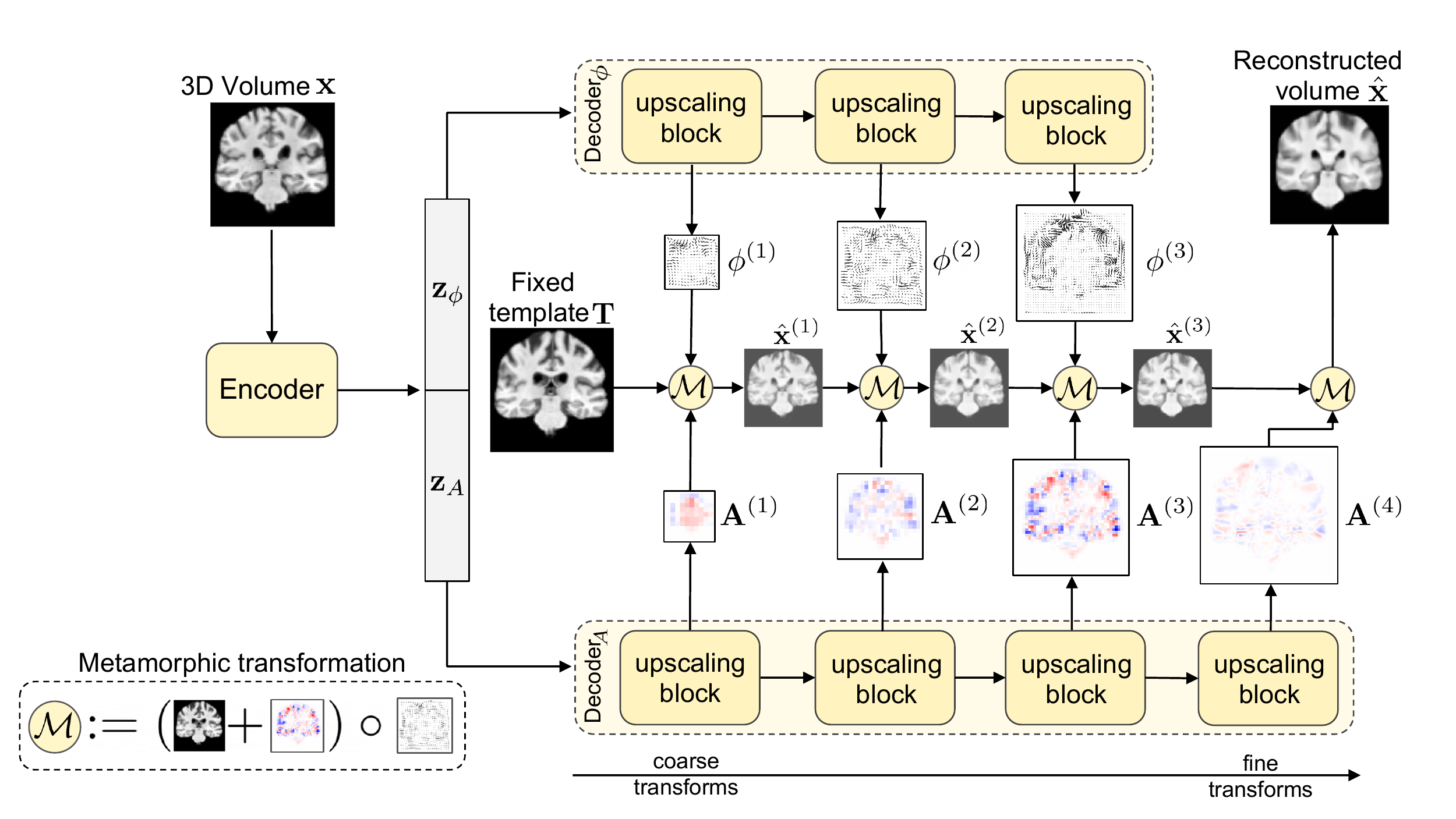}
  \end{subfigure}
  \vspace{-5pt}
  \caption{Description of the proposed \textit{M$^3$AE} model.
    The decoder consists of two separate backbones Decoder$_\phi$ and Decoder$_A$ that output diffeomorphic deformation fields and additive intensity transformations, respectively.
    These transformations are composed from coarse-to-fine scales that increase by a factor of 2 at each level.
    The last transformation only consists of an intensity transformation.
    \vspace{-10pt}
  }
  \label{fig:method}
\end{figure}

In the VAE paradigm, we aim to generate samples, denoted by $\hat{\B x} \in \mathcal{X}$ from a data distribution $p_{\text{data}}(\B x)$ by sampling a random latent variable $\B z \sim \mathcal{N}(\B 0, \B I)$, which is then passed to a decoder to output $p(\B x \vert \B z)$.
During training, given a datapoint $\B x$, we infer the posterior distribution $q(\B z\vert \B x) = \mathcal{N}(\mu_{\B z}, \sigma_{\B z})$ of latent variable $\B z$ through an encoder.
The encoder- and decoder parameters $\theta$ are then jointly optimized to maximize the Evidence Lower BOund (ELBO), which is a lower bound on $\log p(\B x)$ and can be written as
$\T{ELBO}(\theta\vert \B x) = \mathbb{E}_{\B z\sim q(\B z\vert \B x)} \left[ \log p(\B x \vert \B z) \right] + \T{KL}\left[ q(\B z\vert \B x) \Vert \mathcal{N}(\B 0, \B I) \right]$.
This optimization reduces to minimizing a reconstruction term in the data space and a regularization term over the latent space.

The proposed model, which we term \textit{multiscale metamorphic autoencoder (M$^3$AE)} is based on the standard VAE.
However, instead of direct pixelwise outputs, our model outputs morphological transforms that map a fixed standard template $\B {T} \in \mathcal{X}$ to the input image $\B x$. If the template image $\B T$ is sufficiently similar to $\B x$, the generation task may be substantially simpler than generating images `from scratch'.
We use two classes of transforms: deformations and additions.
The diffeomorphic transform $\B T \rightarrow (\B T \circ \B \phi)$, where the deformation field $\phi$ is applied as in the LDDMM framework \cite{ASHBURNER-lddmm}, performs non-affine elastic deformations, which captures atrophy patterns and structural variations.
The intensity transformation is an additive transformation $\B T \rightarrow (\B T + \B A)$, and is expected to capture subject and scanner-specific intensity variations, lesions, and topological irregularities.
The two transforms can be composed into \textit{metamorphic} transformations \cite{metamorphic, metamorphosis}, denoted as $\mathcal{M}(\B T \vert \B \phi, \B A) = (\B T + \B A) \circ \B \phi$.

We then apply these transforms in a cascaded, multiscale fashion at increasing scales. For generation, the decoder is split into two backbones --
Decoder$_{\B \phi}$ and Decoder$_{\B A}$
(Fig. \ref{fig:method}). Each decoder, made up of upscaling blocks, is given a subset of $\B z$ as input (i.e., $\B z_\phi$ / $\B z_A$ to  Decoder$_{\B \phi}$ / Decoder$_{\B A}$, respectively) and outputs a corresponding set of coarse-to-fine transform parameters $\{\B A^{(i)}, \B \phi^{(i)}\}_i$.
By composing coarse and fine transforms in an interleaved fashion as illustrated in Fig.\,\ref{fig:method}, we obtain more expressive transformations and hence better coverage of the data space. The last transformation is only additive in nature and outputs the final reconstruction $\hat{\B x}$.
The loss function, $\mathcal{L}_{\T{total}}$, for training the model can be written as
\begin{align}
  \hspace{0em} \mathcal{L}_{\T{total}}          & = \mathcal{L}_{\T{ELBO}} + \sum\limits_{i=1}^{\T{levels}}\mathcal{L}^{(i)}_{\T{RegRecon}}\label{eq1}                                                                                                                                                                                                                                                       \\
  \hspace{0em} \mathcal{L}_{\T{ELBO}}           & = \Vert \B{x} - \B{\hat{x}}\Vert_1 +  \beta \T{KL}\left[\mathcal{N}(\B{\mu_z}, \B{\sigma_z})\Vert \mathcal{N}(\B 0, \B I)\right]                                                                                                                                                                                                                           \\
  \hspace{0em} \mathcal{L}^{(i)}_{\T{RegRecon}} & = \gamma^{(i)}_1 \Vert \B x - \hat{\B x}^{(i)}\Vert_1 + \underbrace{\gamma_2^{(i)} \Vert \B A^{(i)} \Vert^2 + \gamma_3^{(i)} \Vert \nabla \B \phi^{(i)}\Vert^2 + \gamma_4^{(i)} \Vert \nabla \cdot \B \phi^{(i)}\Vert^2 +   \gamma_5^{(i)} \Vert \B \phi^{(i)}\Vert^2}_{\T{\makebox[0pt]{regularizing coarse-to-fine transform parameters}}}, \label{eq:3}
\end{align}
In Eq. \ref{eq1}, $\mathcal{L}_{\T{ELBO}}$ is the $\beta$VAE \cite{betavae} loss, and $\mathcal{L}^{(i)}_{\T{RegRecon}}$ consists of intermediate regularization and reconstruction loss terms for the $i$th transform level.
To ensure appropriate behaviour of the transformation cascade, we constrain intermediate volumes to be close to the final volume at each level (first term in Eq.\,\ref{eq:3}).
The transformation parameters are regularized using decay terms for $\B A^{(i)}, \B \phi^{(i)}$, and spatial gradient and divergence minimizers  for $\B \phi^{(i)}$. $\beta, \gamma^{(i)}_j$ are hyperparameters to appropriately scale the loss terms.

\begin{table}
  \centering
  \begin{tabular}{ l | l | l | l | l | l | l }
    \multirow{2}{*}{Model}         & \multicolumn{3}{c|}{FID scores} & \multicolumn{3}{c}{reconstruction metrics}                                                                                      \\\cline{2-7}
                                   & axial($\downarrow$)             & coronal($\downarrow$)                      & saggital($\downarrow$) & $L_2$($\downarrow$) & SSIM($\uparrow$) & PSNR($\uparrow$) \\\hline
    $\alpha$WGAN \cite{3dbraingen} & 110.4                           & ${95.4}$                                   & 98.7                   & 0.137               & 0.576            & 12.21            \\
    $\beta$VAE \cite{betavae}      & 178.1                           & 169.5                                      & 175.1                  & $\mathbf{0.045}$    & $\mathbf{0.768}$ & $\mathbf{16.99}$ \\
    $\text{M}^3\text{AE}$ (ours)   & $\mathbf{82.2}$                 & $\mathbf{83.1}$                            & $\mathbf{76.1}$        & 0.047               & 0.759            & 16.81            \\\cline{1-1}\vspace{3pt}
    \grey{Validation set}          & \grey{11.3}                     & \grey{11.5}                                & \grey{8.5}             & \grey{--}           & \grey{--}        & \grey{--}
  \end{tabular}
  \caption{Evaluations of Frechet Inception Distance (FID) scores and reconstruction metrics on the test set. FID scores between the validation set and test set are provided as lower bounds. Arrows indicate whether lower ($\downarrow$) or higher ($\uparrow$) values are better.\vspace{-23pt}}
  \label{table:comparison}
\end{table}

\section{Evaluation}

We evaluated the M$^3$AE model on 3D T1 MRI volumes from the ADNI \cite{adni} dataset.
We used 4789 volumes from a total of 992 unique subjects for training.
An additional 2592 volumes from 526 unique subjects served as the test set.
Volumes were preprocessed by applying bias correction, skull stripping, and linear registration using FSL\footnote{\href{https://fsl.fmrib.ox.ac.uk/fsl/fslwiki}{FMRIB Software Library v6.0
    Created by the Analysis Group, FMRIB, Oxford, UK.}}, and were then downscaled and cropped to $80\times96\times80$ pixels, mainly to reduce training times for the baseline and the proposed method. We used the congitively normal baseline scan for subject \texttt{003S692} as the fixed template.

The model's $\T{Encoder},\T{Decoder}_{\B \phi}$ and $\T{Decoder}_{\B A}$ are made up of 3D convolutions.
We used four layers of morphological transforms, where each layer progressively doubles in each dimension until it reaches the original volume size.
The model was trained using the AdamW optimizer with a learning rate of 0.0003.
As our first baseline we chose the $\alpha$WGAN \cite{3dbraingen}, which consists of an autoencoder with adversarial losses over the generated volume and the latent space vector. We retrained the model using the code provided by the authors. We also evaluated against a standard $\beta$VAE \cite{betavae}. For $\beta$VAE as well as for our proposed M$^3$AE we set $\beta=3$.
In future work, we also aim to include the 3D StyleGAN \cite{3d-stylegan} in our evaluations. Unfortunately, we could not include it in this submission due to implementation issues and time constraints.

We quantitatively assessed the generation quality of M$^3$AE with the $\alpha$WGAN baseline using FID \cite{fid} scores on 2D slices of randomly generated volumes.
Following related work \cite{3d-stylegan}, we chose four slices (at locations [30,40,50,60] of the volume) for each of the three axes and averaged the per-axis FID scores over them. We also computed reconstruction metrics in the form of SSIM, PSNR, and MSE.

Table \ref{table:comparison} shows the values obtained for the above metrics.
M$^3$AE substantially outperformed $\alpha$WGAN as well as $\beta$VAE in FID scores. %
The proposed model also outperformed $\alpha$WGAN in terms of reconstruction quality.
$\beta$VAE and M$^3$AE perform similarly in terms of reconstruction metrics although the $\beta$VAE has a slight advantage. This can be explained by the morphological constraint placed on our model which limit outputs to transformations of the fixed template. This constraint, however, is also what enables the superior FID scores. %

Qualitative analysis of the compared methods reveals that samples from $\beta$VAE are not anatomically correct in many cases and display regions that appear scrambled.
$\alpha$WGAN generates anatomically viable samples, however in many of them the cortical folds do not follow the anatomical structure. Additionally, as in $\beta$VAE, samples exhibit regions where artifacts are dramatically different from the rest of the volume.
Samples from our model are the most anatomically correct due to starting from the fixed template. However, the samples occasionally have a wavy visual quality due to improperly generated random deformation fields at finer scales. We furthermore observe a lack of sufficient topological diversity in the cortical folds.
Samples for each of the above methods are shown in Section \ref{sec:samples} of Supplementary Material.

\vspace{-5pt}
\section{Discussion}
\vspace{-5pt}
This work presents a novel generative modeling approach for 3D MRI synthesis taking advantage of strong anatomical priors and multiscale morphological transformations.
Quantitative analysis shows that our model obtains substantially better coverage of the data distribution as well as comparable or better reconstruction quality compared to baselines. In contrast to GAN based approaches our model also retains an expressive latent space which enables further downstream analysis.
Our initial results show that using a fixed template to take advantage of anatomical knowledge offers a promising avenue for MRI volume synthesis.
In future works, we aim to address current limitations and apply this model for longitudinal modeling of brain MRIs in the context of Alzheimer's disease.

\section{Potential negative impact}
This work aims to generate high-fidelity MRI volumes for better downstream analyses on a range of domain-specific tasks in medical imaging and clinical research. If we do not train the model on data representative of the clinical population, it may introduce biases in the downstream analyses and thus lead to sub-optimal or harmful predictions. This may also happen if the model does not equitably model features in the data based on protected attributes.

\section{Acknowledgements}
This work is part of a project funded by the Deutsche Forschungsgemeinschaft (DFG, German Research Foundation) under Germany’s Excellence Strategy – EXC number 2064/1 – Project number 390727645.
This work was supported by the German Federal Ministry of Education and Research (BMBF): Tübingen AI Center, FKZ: 01IS18039A.
The authors thank the International Max Planck Research School for Intelligent Systems (IMPRS-IS) for supporting Jaivardhan Kapoor.
The authors also thank Dr. Sergios Gatidis (MPI-IS, Germany) for helpful feedback.

\printbibliography

\appendix

\begin{center}
  \Large{\textbf{Supplementary Material}}
\end{center}

\section{Generated Samples}
\label{sec:samples}

\begin{figure}[ht!]
  \centering
  \begin{subfigure}[b]{0.99\textwidth}
    \centering
    \includegraphics[width=0.85\textwidth]{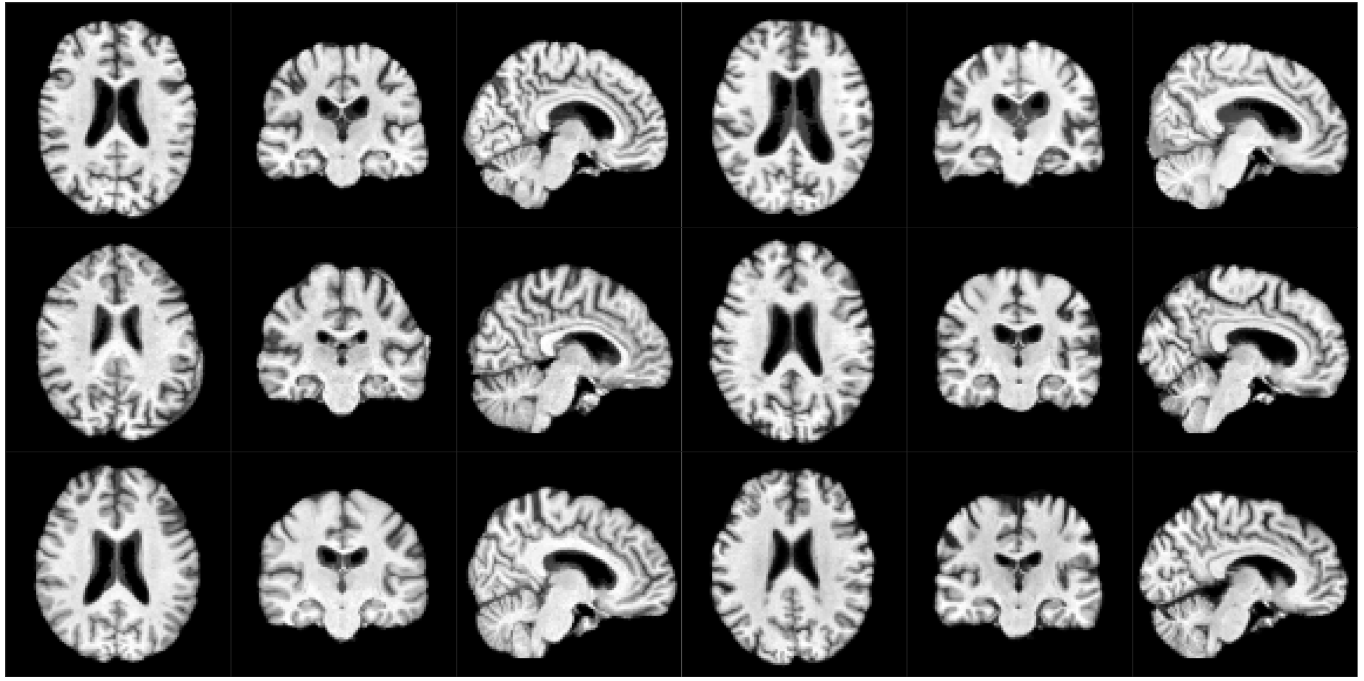}
  \end{subfigure}
  \caption{
    Preprocessed volumes from the ADNI dataset
  }
  \label{fig:real_images}
\end{figure}

\begin{figure}[ht!]
  \centering
  \begin{subfigure}[b]{0.99\textwidth}
    \centering
    \includegraphics[width=0.85\textwidth]{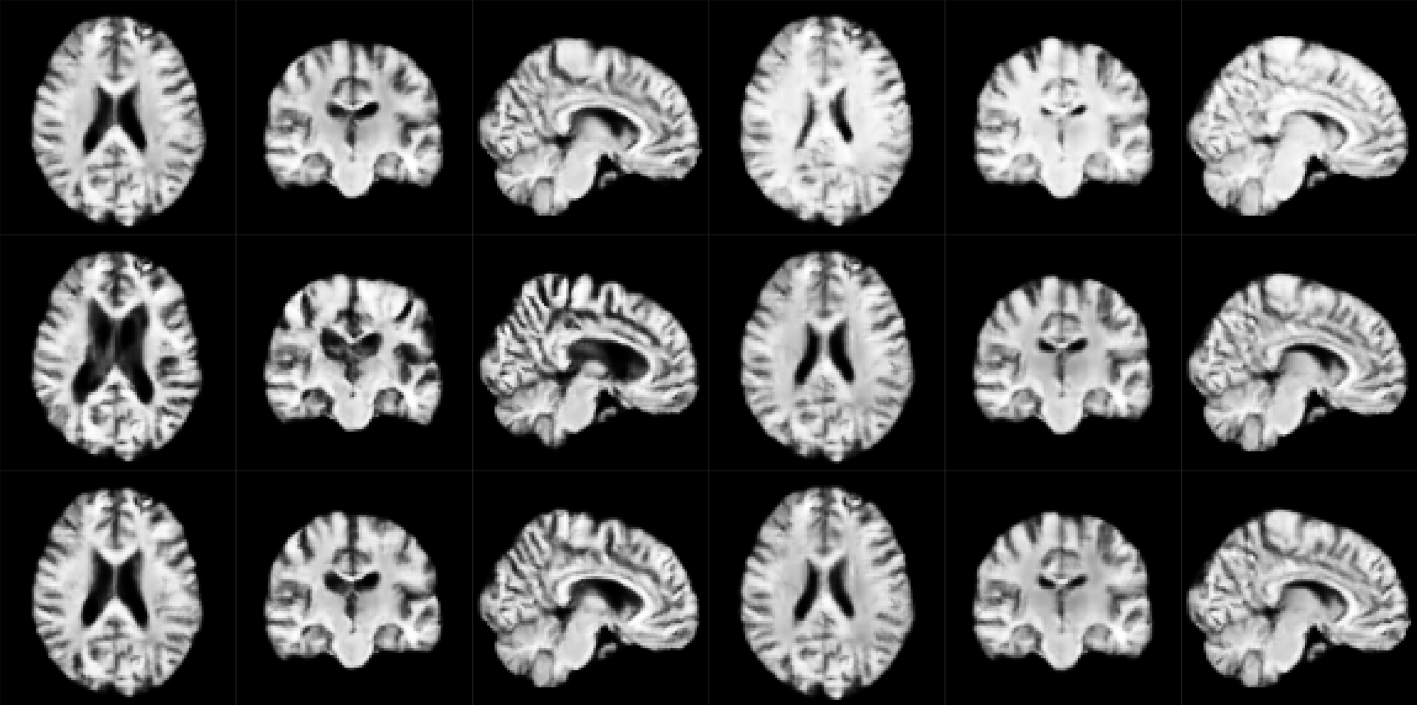}
  \end{subfigure}
  \caption{
    Samples from $\alpha$WGAN \cite{3dbraingen}
  }
  \label{fig:samples_alphawgan}
\end{figure}

\begin{figure}[ht!]
  \begin{subfigure}[b]{0.99\textwidth}
    \centering
    \includegraphics[width=0.85\textwidth]{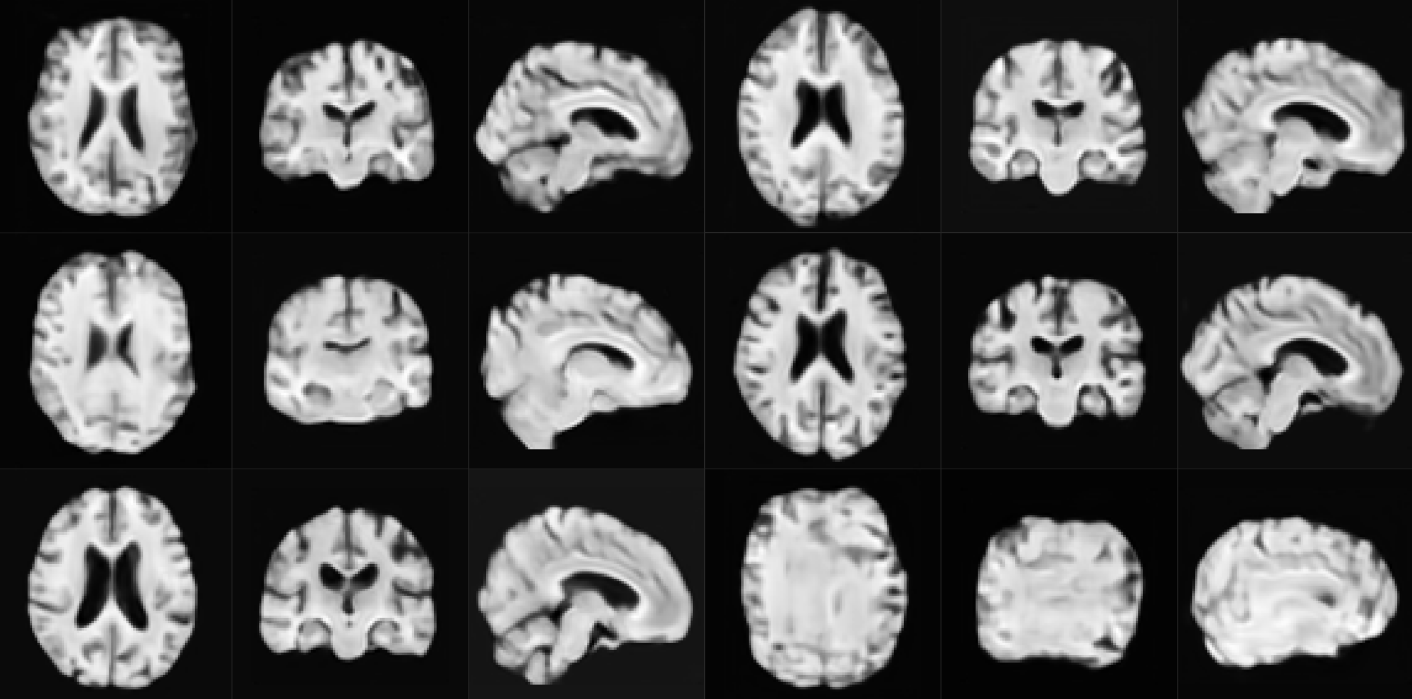}
  \end{subfigure}
  \caption{
    Samples from $\beta$VAE \cite{betavae}
  }
  \label{fig:samples_vae}
\end{figure}

\begin{figure}[ht!]
  \centering
  \begin{subfigure}[b]{0.99\textwidth}
    \centering
    \includegraphics[width=0.85\textwidth]{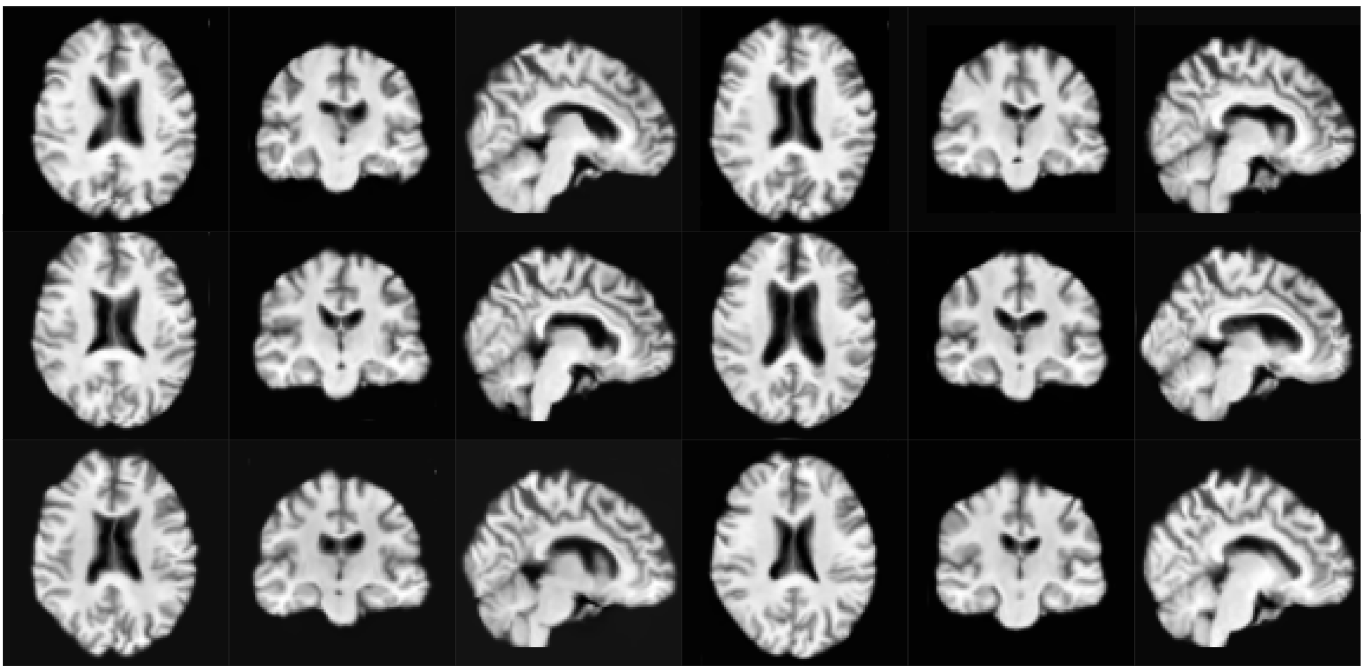}
  \end{subfigure}
  \caption{
    Samples from proposed model M$^3$AE
  }
  \label{fig:samples_m3ae}
\end{figure}

\section{M$^3$AE architecture and additional training details}

\begin{figure}[ht!]
  \centering
  \begin{subfigure}[b]{0.99\textwidth}
    \centering
    \includegraphics[width=0.85\textwidth]{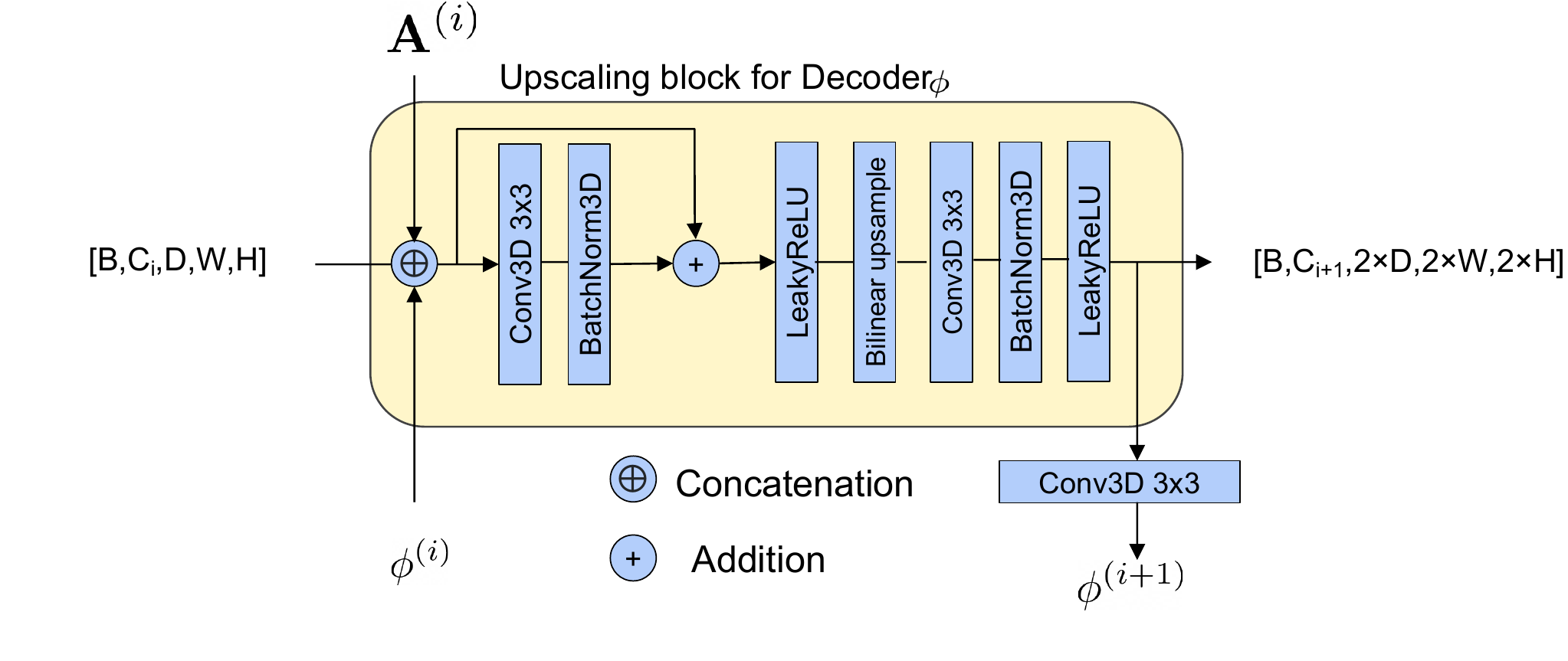}
  \end{subfigure}
  \caption{
    Upsampling block architecture for M$^3$AE
  }
  \label{fig:upsampleblock}
\end{figure}

The encoder and decoder of the M$^3$AE are made of downsampling and upsampling blocks respectively. A schematic of the upsampling block is show in Fig. \ref{fig:upsampleblock}. The downsampling block uses the same architecture, with the 2x bilinear upscaling operation replaced by downscaling. The upsampling block receives the utput of the previous block concatenated with the previous  block's transformation outputs. The block outputs the corresponding transformation by passing through a 3d convolution layer, which reduces the number of channels to three (in the case of diffeomorphic transform) or one (in the case of additive transform).

We used a fully connected latent space with a size of 512. We split it into 2 vectors of size 256 each, which were reshaped to a shape of [128,5,6,5] and then passed through the upscaling blocks of the respective decoders. The number of hidden channels were set to [128, 64, 32, 16].

The same hidden channel configuration was set in reverse for the downscaling blocks in the encoder, where the final output of shape [128, 5,6,5] was flattened and resized through a fully connected layer to output the 512-dimension mean and log variance of latent vector.

The model was trained for a total of 30 epochs. We used the following values for the loss scaling terms: $\beta=6.0,\gamma_1^{(1)}=0.25, \gamma_1^{(2)}=0.25, \gamma_1^{(3)}=0.5, \gamma_2^{(1)}=0.6, \gamma_2^{(2)}=1.2, \gamma_2^{(3)}=1.2,\gamma_2^{(4)}=2.4, \gamma_3^{(1)}=0.6, \gamma_3^{(2)}=1.2, \gamma_3^{(3)}=1.2,\gamma_4^{(1)}=0.6, \gamma_4^{(2)}=1.2, \gamma_4^{(3)}=1.2,\gamma_5^{(1)}=0.6, \gamma_5^{(2)}=1.2, \gamma_5^{(3)}=1.2$.
These were chosen using a hyperparameter search on the validation set.
During training, the KL divergence loss weighing term $\beta$ was slowly increased from 0 to 6 across 10 epochs.

\end{document}